\documentclass[prl,twocolumn,10pt,letterpaper,tightenlines]{revtex4}%
\usepackage{amsmath}
\usepackage{graphicx}%
\usepackage{amsfonts}%
\usepackage{amssymb}

\begin{document}

\title{\textbf{Generalized Cooper Pairing and Bose-Einstein Condensation}}
\author{V.C. Aguilera-Navarro$^{a}$, M. Fortes$^{b}$, and M. de Llano$^{c}$}
\affiliation{$^{a}$Depot. de Qu\'{\i}mica e F\'{\i}sica,
UNICENTRO, 85015, Guarapuava, PR, Brazil}
\affiliation{$^{b}$Instituto de F\'{\i}sica, Universidad Nacional
Aut\'{o}noma de M\'{e}xico, Apdo. Postal 20-364, 01000 M\'{e}xico,
D.F., Mexico} \affiliation{$^{c}$Instituto de Investigaciones en
Materiales, Universidad Nacional Aut\'{o}noma de M\'{e}xico, Apdo.
Postal 70-360, 04510 M\'{e}xico, DF, Mexico}

\begin{abstract}
A Bethe-Salpeter treatment of Cooper pairs (CPs) based on an ideal Fermi gas
(IFG) ``sea''\ yields the familiar negative-energy, two-particle bound-state
if two-hole CPs are ignored, but is meaningless otherwise as it gives
purely-imaginary energies. However, when based on the BCS ground state,
legitimate two-particle ``moving''\ CPs emerge but as positive-energy,
finite-lifetime resonances for nonzero center-of-mass momentum, with a
\textit{linear }dispersion leading term. Bose-Einstein condensation of such
pairs may thus occur in exactly two dimensions as it cannot with quadratic dispersion.

\textbf{PACS }05.30.Fk;\ 05.30.Jp; 71.10.-w; 74.20.Fg\

\end{abstract}
\maketitle

Shortly after the publication of the BCS theory \cite{BCS}\ of
superconductivity charged Cooper pairs \cite{Coo} (CPs) observed in magnetic
flux quantization experiments with 3D conventional \cite{classical}%
\cite{classical2}, and much later with quasi-2D cuprate
\cite{cuprates} superconductors, suggested CPs as an indispensable
ingredient. Although BCS theory admits the presence of Cooper
``correlations,''\ several boson-fermion (BF) models
\cite{BF4}-\cite{BF10} with real, bosonic CPs have been introduced
after the pioneering work of Refs. \cite{BF}-\cite{BF2}. However,
with one exception \cite{BF7a}-\cite{CMT02}, all such models
neglect the effect of two-hole (2h) CPs treated on an equal
footing with two-particle (2p) CPs---as Green's functions
\cite{FW}\ can naturally guarantee.

The BCS condensate consists of equal numbers of 2p and 2h Cooper
correlations; this is evident from the perfect symmetry\ about
$\mu$, the electron chemical potential, of the well-known
Bogoliubov \cite{Bog} $v^{2}(\epsilon)$\ and $u^{2}(\epsilon)$
coefficients [see just below (\ref{mCP}) later on], where
$\epsilon$ is the electron energy.\ Some motivation for this
Letter comes from the unique but unexplained role played by
\textit{hole }charge carriers in the normal state of
superconductors in general \cite{Hirsch}, as well as from the
ability of the ``complete (in that both 2h- and 2p-CPs are allowed
in varying proportions) BF model'' of Refs.
\cite{BF7a}-\cite{CMT02} to ``unify'' both BCS and Bose-Einstein
condensation (BEC) theories as special cases. Substantially higher
$T_{c}$'s than BCS theory are then predicted without abandoning
electron-phonon dynamics. Compelling evidence for a significant
presence of this dynamics in high-$T_{c}$ cuprate superconductors
from angle-resolved photoemission spectroscopy data has recently
been reported \cite{Shen}.

In this Letter the Bethe-Salpeter (BS)
many-body equation (in the ladder approximation)\ treating both 2p
and 2h pairs on an equal footing is used to show that, while the
ordinary CP problem [based on an ideal Fermi gas (IFG) ground
state (the usual ``Fermi sea'')] does \textit{not} possess stable
energy solutions: i) CPs based not on the IFG-sea but on the BCS
ground state survive as \textit{positive} energy resonances; ii)
their dispersion relation in leading order in the total (or
center-of-mass) momentum (CMM)
$\hbar\mathbf{K\equiv\hbar(k}_{1}+\mathbf{k}_{2})$ is
\textit{linear }rather than the quadratic $\hbar^{2}K^{2}/4m$ of a
composite boson (e.g., a deuteron) of mass $2m$\ moving not in the
Fermi sea but in vacuum; and iii) this latter ``moving CP''
solution, though often confused with it, is physically
\textit{distinct }from another more common solution sometimes
called the
Anderson-Bogoliubov-Higgs (ABH) \cite{ABH}, (\cite{BTS} p. 44), \cite{Higgs}%
-\cite{Traven2} collective excitation. The ABH mode is also linear in leading
order and goes over into the IFG ordinary sound mode in zero coupling. A new
feature emerging from our present 2D results, compared with a prior 3D study
outlined in Ref. \cite{Honolulu}, is the imaginary energy term leading to
finite-lifetime CPs. We focus here on 2D because of its interest
\cite{Varma}\cite{Sachdev}\ for quasi-2D cuprate superconductors. In general,
our results will be crucial for Bose-Einstein condensation (BEC) scenarios
employing BF models of superconductivity, not only \textit{in} \textit{exactly
2D} as with the Berezinskii-Kosterlitz-Thouless \cite{BKT}\cite{KT}
transition, but also down to ($1+\epsilon$)D which characterize the quasi-1D
organo-metallic (Bechgaard salt) superconductors \cite{organometallics}%
-\cite{jerome2}. Striking experimental confirmation of how
superconductivity is ``extinguished'' as dimensionality $d$ is
diminished towards unity has been reported by Tinkham and
co-workers \cite{Tinkham01}\cite{Tinkham00}. They measured
resistance vs. temperature curves in superconducting nanowires
consisting of carbon nanotubes sputtered with amorphous
$Mo_{79}Ge_{21}$ and of widths from 22 to 10 nm, showing how
$T_{c}$ vanishes for the thinnest widths.
Our results also apply,
albeit with a different interaction, to neutral-atom superfluidity
as in liquid $^{3}$He\ \cite{He3} as well as to ultracold
trapped alkali Fermi gases such as $^{6}$Li \cite{Li6}\ and $^{40}$K \cite%
{Holland} since pairing is believed to occur there also.

For bosons with excitation energy $\varepsilon_{K}=C_{s}K^{s}+o(K^{s})$ (for
small CMM $K$) BEC occurs in a box of length $L$\ if and only if $d>s,$ since
$T_{c}\equiv0$ for all $d\leq s$. The commonest example is $s=2$ as in the
textbook case of ordinary bosons with $\varepsilon_{K}=$ $\hbar^{2}K^{2}/2m$
exactly, giving the familiar result that BEC is not allowed for $d\leq2$. The
general result for any $s$ is seen as follows. The total boson number is
\[
N=N_{0}(T)+\sum_{\mathbf{K\neq0}}[\exp\beta(\varepsilon_{\mathbf{K}}-\mu
_{B})-1]^{-1}%
\]
with $\beta\equiv k_{B}T$. Since $N_{0}(T_{c})\simeq0$ while the boson
chemical potential $\mu_{B}$ also vanishes\ at $T=T_{c}$, in the thermodynamic
limit the boson number density becomes%
\[
N/L^{d}\simeq A_{d}\int_{0^{+}}^{\infty}\mathrm{d}KK^{d-1}[\exp\beta_{c}%
(C_{s}K^{s}+\cdots)-1]^{-1}%
\]
where $A_{d}$ is a finite coefficient. Thus%
\[
N/L^{d}\simeq A_{d}(k_{B}T_{c}/C_{s})\int_{0^{+}}^{K_{\max}}\mathrm{d}%
KK^{d-s-1}+\int_{K_{\max}}^{\infty}\cdots,
\]
where $K_{\max}$ is small and can be picked arbitrarily so long as the
integral $\int_{K_{\max}}^{\infty}\cdots$ is finite, as is $N/L^{d}$. However,
if $d=s$ the first integral\ gives $\ln K\mid_{_{0}}^{K_{\max}}=-\infty$; and
if $d<s$ it gives $1/(d-s)K^{s-d}\mid_{_{0}}^{K_{\max}}=-\infty$. Hence,
$T_{c}${\small \ }must vanish{\small \ }if and only if{\small \ }$d\leq
s$,{\small \ }but is otherwise finite. This conclusion hinges \textit{only }on
the leading term of the boson dispersion relation $\varepsilon_{K}$.\ The case
$s=1$ emerges in the CP problem to be discussed now.

In dealing with the many-electron system we assume a BCS-like electron-phonon
model $s$-wave inter-electron interaction, whose double Fourier
transform\ $\nu(|\mathbf{k}_{1}-\mathbf{k}_{1}^{\prime}|)$ is just
\begin{equation}
\nu(k_{1},k_{1}^{\prime})=-(k_{F}/k_{1}^{\prime})V \label{int}%
\end{equation}
if$\;k_{F}-k_{D}<k_{1}<k_{F}+k_{D}$, and \ $=0$\ otherwise. Here $V>0$, $\hbar
k_{F}\equiv mv_{F}$ the Fermi momentum, $m$ the effective electron mass,
$v_{F}$ the Fermi velocity, and $k_{D}\equiv\omega_{D}/v_{F}$ with $\omega
_{D}$ the Debye frequency. The usual condition $\hbar\omega_{D}\ll E_{F}$ then
implies that $k_{D}/k_{F}\equiv\hbar\omega_{D}/2E_{F}\ll1$.

The BS wavefunction equation \cite{Honolulu} in the ladder approximation\ with
both particles and holes for the original IFG-based CP problem using
(\ref{int}) leads to an equation for the wavefunction $\psi_{\mathbf{k}}$ in
momentum space for CPs with \textit{zero}\ CMM $\mathbf{K\equiv k}%
_{1}+\mathbf{k}_{2}=0$ that is
\begin{equation}
(2\xi_{k}-\mathcal{E}_{0})\psi_{\mathbf{k}}=V\sum_{\mathbf{k}^{\prime}}%
{}^{^{\prime}}\psi_{\mathbf{k}^{\prime}}-V\sum_{\mathbf{k}^{\prime}}%
{}^{^{\prime\prime}}\psi_{\mathbf{k}^{\prime}}.\label{CP}%
\end{equation}
Here $\xi_{k}\equiv\hbar^{2}k^{2}/2m-E_{F}$, $\mathcal{E}_{0}$ is the
eigenvalue energy and $\mathbf{k\equiv%
\frac12
({k}_{1}-{k}_{2})}$ is the relative wavevector of a pair. The single prime
over the first (2p-CP) summation term denotes the restriction $0<\xi
_{k^{\prime}}<\hbar\omega_{D}$ while the double prime in the last (2h-CP) term
means $-\hbar\omega_{D}<\xi_{k^{\prime}}<0$. Without this latter term we have
Cooper's Schr\"{o}dinger-like equation \cite{Coo}\ for 2p-CPs whose implicit
solution is clearly $\psi_{\mathbf{k}}=(2\xi_{k}-\mathcal{E}_{0})^{-1}%
V\sum_{\mathbf{k}^{\prime}}^{^{\prime}}\psi_{\mathbf{k}^{\prime}}.$ Since the
summation term is constant, performing that summation on both sides allows
canceling the $\psi_{\mathbf{k}}$-dependent terms, leaving the eigenvalue
equation $\sum_{\mathbf{k}}^{^{\prime}}(2\xi_{k}-\mathcal{E}_{0})^{-1}%
=1/V$\ with the familiar solution $\mathcal{E}_{0}=-2\hbar\omega
_{D}/(e^{2/\lambda}-1)$\ (exact in 2D, and to a very good approximation
otherwise if $\hbar\omega_{D}\ll E_{F}$) where $\lambda\equiv VN(E_{F})$ with
$N(E_{F})$ the electronic density of states (DOS) for one spin. This
corresponds to a negative-energy, stationary-state bound pair. For
$K\geqslant0$ the CP eigenvalue equation becomes
\begin{equation}
\sum_{\mathbf{k}}{}^{^{\prime}}(2\xi_{k}+\hbar^{2}K^{2}/2m-\mathcal{E}%
_{K})^{-1}=1/V.\label{CPKeqn}%
\end{equation}
Note that a CP state of energy $\mathcal{E}_{K}$\ is characterized only by a
definite $\mathbf{K}$ but \textit{not }definite $\mathbf{k}$, in contrast to a
``BCS pair'' defined \cite{BCS}\ with fixed\ $\mathbf{K}$ and $\mathbf{k}$ (or
equivalently definite $\mathbf{k}_{1}$\ and $\mathbf{k}_{2}$). Without the
first summation term in (\ref{CP})\ the same result in $\mathcal{E}_{0}$\ for
2p-CPs follows for 2h-CPs (apart from a sign change). However, using similar
techniques to solve the \textit{complete }equation (\ref{CP})---which
\textit{cannot }be derived from an ordinary (non-BS) Schr\"{o}dinger-like
equation in spite of its simple appearance---gives the purely-imaginary
$\mathcal{E}_{0}=\pm i2\hbar\omega_{D}/\sqrt{e^{2/\lambda}-1}$, thus implying
an obvious instability. This was reported in Refs. \cite{BTS} p. 44 and
\cite{AGD} who did not stress the pure 2p and 2h cases just discussed. Clearly
then, the original CP picture is\textit{ meaningless }if particle- and
hole-pairs are treated on an equal footing as consistency demands. This is
perhaps the prime motivation for seeking a new unperturbed Hamiltonian about
which to, e.g., do perturbation theory.

A BS treatment not about the IFG sea but about the BCS ground
state \textit{vindicates the CP concept}. This substitution might
seem an artificial mathematical construct but its experimental
support lies precisely in Refs. \cite{classical}-\cite{cuprates}
and its physical justification lies in recovering two expected
results: the ABH sound mode as well as finite-lifetime effects in
CPs. In either 3D \cite{Honolulu}\ or 2D the BS equation yields
two \textit{distinct} solutions: the usual ABH sound solution and
a highly nontrivial ``moving CP'' solution. The BS formalism gives
rise to a set of three coupled equations, one for each (2p, 2h and
ph) channel\ wavefunction\ for any spin-independent interaction
such as (\ref{int}). However, the ph channel decouples, leaving
only two coupled wavefunction\ equations for the ABH solution. The
equations involved are too lengthy, and will be derived in detail
elsewhere. The \textit{ABH collective excitation mode }energy
$\mathcal{E}_{K}$\ is found to be determined by an equation that
for $\mathbf{K}=0$ gives $\mathcal{E}_{0}=0$ (Ref. \cite{BTS} p.
39) and reduces to $\int_{0}^{\hbar\omega_{D}}d\xi/\sqrt{\xi^{2}+\Delta^{2}%
}=1/\lambda$,$\ $the familiar BCS $T=0$ gap equation for interaction
(\ref{int}) whose solution is $\Delta=$\ $\hbar\omega_{D}/\sinh(1/\lambda)$.
Taylor-expanding $\mathcal{E}_{K}$ about $K=0$ and small $\lambda$ gives%
\begin{subequations}
\begin{equation}
\mathcal{E}_{K}\simeq\frac{\hbar v_{F}}{\sqrt{2}}K+O(K^{2}).\tag{4}%
\end{equation}
Note that the leading term is just the ordinary sound mode in an IFG whose
sound speed $c$ $=$ $v_{F}/\sqrt{d}$ in $d$ dimensions which also follows
trivially from the zero-temperature IFG pressure $P=n^{2}[d(E/N)/dn]=2nE_{F}%
/(d+2)$ on applying the familiar thermodynamic relation $dP/dn=mc^{2}$. Here
$E=dE_{F}/(d+2)$ is the IFG ground-state energy while $n\equiv N/L^{d}%
=k_{F}^{d}/d2^{d-2}\pi^{d/2}\Gamma(d/2)$ the fermion-number density.

The second solution in the BCS-ground-state-based BS treatment is the
\textit{moving CP }solution for the pair energy $\mathcal{E}_{K}$ which in 2D
is contained in the equation
\end{subequations}
\begin{gather}
\frac{1}{2\pi}\lambda\hbar v_{F}\int_{k_{F}-k_{D}}^{k_{F}+k_{D}}dk\int
_{0}^{2\pi}d\varphi u_{\mathbf{K}/2+\mathbf{k}}v_{\mathbf{K}/2-\mathbf{k}%
}\times\nonumber\\
\times\{u_{\mathbf{K}/2-\mathbf{k}}v_{\mathbf{K}/2+\mathbf{k}}-u_{\mathbf{K}%
/2+\mathbf{k}}v_{\mathbf{K}/2-\mathbf{k}}\}\times\nonumber\\
\times\frac{E_{\mathbf{K}/2+\mathbf{k}}+E_{\mathbf{K}/2-\mathbf{k}}%
}{-\mathcal{E}_{K}^{2}+(E_{\mathbf{K}/2+\mathbf{k}}+E_{\mathbf{K}%
/2-\mathbf{k}})^{2}}=1, \label{mCP}%
\end{gather}
where $\varphi$ is the angle between $\mathbf{K}$ and $\mathbf{k}$;
$\lambda\equiv VN(E_{F})$ as before with $N(E_{F})\equiv m/2\pi\hbar^{2}$ the
constant 2D DOS and $V$ the interaction strength defined in (\ref{int});
$E_{\mathbf{k}}\equiv\sqrt{\xi_{k}{}^{2}+\Delta^{2}}$ with $\Delta$ the
fermionic gap; while $u_{k}^{2}\equiv%
\frac12
(1+\xi_{k}/E_{\mathbf{k}})$ and $v_{k}^{2}\equiv1-u_{k}^{2}$ are the
Bogoliubov functions \cite{Bog}. In addition to the pp and hh wavefunctions
(depicted grafically in Ref. \cite{Honolulu} Fig. 2), diagrams associated with
the ph channel give zero contribution at $T=0$. A third equation for the ph
wavefunction describes the ph bound state but turns out to depend only on the
pp and hh wavefunctions. Taylor-expanding $\mathcal{E}_{K}$ in powers of
$K$\ around $K=0$, and introducing a possible damping factor by adding an
imaginary term $-i\Gamma_{K}$ in the denominator, yields to order $K^{2}$ for
small $\lambda$%
\begin{align}
\pm\mathcal{E}_{K}  &  \simeq2\Delta+\frac{\lambda}{2\pi}\hbar v_{F}%
K+\frac{1}{9}\frac{\hbar v_{F}}{k_{D}}e^{1/\lambda}K^{2}\nonumber\\
&  -i\left[  \frac{\lambda}{\pi}\hbar v_{F}K+\frac{1}{12}\frac{\hbar v_{F}%
}{k_{D}}e^{1/\lambda}K^{2}\right]  +O(K^{3}) \label{linquadmCP}%
\end{align}
where the upper and lower sign refers to 2p- and 2h-CPs, respectively. A
linear dispersion in leading order again appears, but now associated with the
bosonic moving CP. The \textit{positive}-energy 2p-CP resonance has a lifetime
$\tau_{K}\equiv$ $\hbar/2\Gamma_{K}=\hbar/2\left[  (\lambda/\pi)\hbar
v_{F}K+(\hbar v_{F}/12k_{D})e^{1/\lambda}K^{2}\right]  $ diverging only at
$K=0$, and falling to zero as $K$ increases. Thus, ``faster'' moving CPs are
shorter-lived and eventually break up, while ``non-moving'' ones are
stationary states. The linear term $(\lambda/2\pi)\hbar v_{F}K$ contrasts
sharply with the \textit{coupling-independent} leading-term in $\mathcal{E}%
_{K}=\mathcal{E}_{0}-(2/\pi)\hbar v_{F}K+O(K^{2})$ (or $1/2$ in 3D
\cite{Schrieffer}\ instead of $2/\pi$) that follows from the \textit{original
}CP problem (\ref{CPKeqn})\ neglecting holes---for either interaction
(\ref{int}) \cite{PhysC98} \textit{or }an attractive delta inter-fermion
potential \cite{PRB2000}\cite{PhysicaC} (imagined regularized \cite{GT} to
have a single bound state whose binding energy serves as the coupling
parameter). In the latter simple example, moreover, it is manifestly clear\ in
2D\ \cite{PRB2000}\ that the quadratic $\hbar^{2}K^{2}/4m$\ stands alone\ as
the leading term for any coupling only when $E_{F}\equiv$ $%
\frac12
mv_{F}^{2}$ is \textit{strictly }zero, i.e., in the absence of the Fermi sea.
Fig. 1 graphs\ the exact moving CP (mCP)\ energy extracted from (\ref{mCP}),
along with its leading linear-dispersion term and this plus the next
(quadratic) term from (\ref{linquadmCP}). The interaction parameter values
used in (\ref{int})\ were $\hbar\omega_{D}/E_{F}=0.05$ (a typical value for
cuprates) and the two values $\lambda=%
\frac14
$ and $%
\frac12
$, giving for $\mathcal{E}_{0}/E_{F}\equiv2\Delta/E_{F}=2\hbar\omega_{D}%
/E_{F}\sinh(1/\lambda)\simeq0.004$ and $0.028,$ respectively (marked as dots
in the figure). Remarkably enough, the linear approximation (thin short-dashed
lines in figure) is better over a wider range of $K/k_{F}$ values for weaker
coupling in spite of a larger and larger partial contribution from the
quadratic term in (\ref{linquadmCP}); this peculiarity also emerged from the
ordinary CP treatment of Ref. \cite{PhysC98} and might suggest the expansion
in powers of $K$ to be an asmyptotic series that should be truncated after the
linear term. For reference we also plot the linear term $\hbar v_{F}K/\sqrt
{2}$\ of the sound solution (4).

We cannot presently address such matters as the nature of the normal state,
the pseudogaps observed in underdoped cuprates, etc., but efforts in these
directions are in progress.%

\begin{figure}
[ptb]
\begin{center}
\includegraphics[
height=2.0583in,
width=3.3313in
]%
{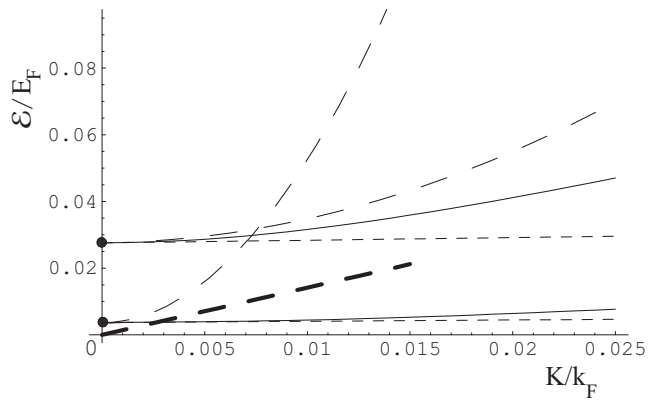}%
\caption{Exact \textquotedblleft moving Cooper pair\textquotedblright\ energy
$\mathcal{E}_{K}$\ (in units of $E_{F}$) from (\ref{mCP})\ (full curves),
compared with its linear leading term (short-dashed lines) and its linear plus
quadratic expansion (long-dashed curves) both from (\ref{linquadmCP}),
\textit{vs} CMM wavenumber $K$ (in units of $k_{F})$, for interaction (1)
parameters $\lambda=\frac14 $ (lower set of curves) and $\frac12 $ (upper set
of curves), and $\hbar\omega_{D}/E_{F}=0.05$. For reference, the leading
linear term (4) of the ABH sound mode is also plotted (lower thick dashed
line).}%
\end{center}
\end{figure}

Like Cooper's \cite{Coo} [see Eq. (\ref{CPKeqn})], our BS CPs are
characterized by a definite $\mathbf{K}$ and \textit{not} also by definite
$\mathbf{k}$ as the pairs discussed by BCS \cite{BCS}. Hence, the objection
does not apply that CPs are not bosons because BCS pairs with
definite\ $\mathbf{K}$ and $\mathbf{k}$ (or equivalently definite
$\mathbf{k}_{1}$\ and $\mathbf{k}_{2}$) have creation/annihilation operators
that do \textit{not }obey Bose commutation relations [Ref. \cite{BCS}, Eqs.
(2.11) to (2.13)]. In fact, either (\ref{CPKeqn}) or (\ref{mCP}) shows that a
given ``ordinary'' or BS CP state labeled by either $\mathbf{K}$ or
$\mathcal{E}_{K}$\ can accommodate (in the thermodynamic limit) an
indefinitely many possible BCS pairs with different $\mathbf{k}$'s. This
implies BE statistics for either ordinary or BS CPs as each energy state has
no occupation limit.

To conclude, hole pairs treated on a par with electron pairs play a vital role
in determining the precise nature of CPs even at zero temperature, only when
based not on the usual ideal-Fermi-gas (IFG) ``sea'' but on the BCS ground
state. Treatment them with the Bethe-Salpeter equation gives
purely-imaginary-energy CPs when based on the IFG, and positive-energy
resonant-state CPs with a finite lifetime for nonzero CMM when based on the
BCS ground state---instead of the more familiar negative-energy stationary
states of the original IFG-based CP problem that neglects holes, as sketched
just below (\ref{CP}). The BS ``moving-CP'' dispersion relation is gapped by
twice the BCS energy gap, followed by a \textit{linear} leading term in the
CMM expansion about $K=0$. This linearity is distinct from the better-known
one associated with the sound or ABH collective excitation mode whose energy
vanishes at $K=0$. Thus, boson-fermion models assuming this CP linearity for
the boson component instead of the quadratic\ $\hbar^{2}K^{2}/4m$ can give BEC
for all $d>1$, including\ exactly 2D, and thus in principle apply not only to
quasi-2D cuprate but also to quasi-1D organo-metallic superconductors.

\bigskip

Partial support is acknowledged from grant IN106401, PAPIIT (Mexico). MdeLl
thanks P.W. Anderson, M. Casas, J.R. Clem, D.M. Eagles, S. Fujita, and K.
Levin for comments, and with MF is grateful to M. Grether, O. Rojo, M.A.
Sol\'{\i}s, V.V. Tolmachev, A.A. Valladares and H. Vucetich for discussions.
Both VCAN and MdeLl thank CNPq (Brazil) and CONACyT (Mexico) for bilateral support.

\end{document}